\documentclass[twocolumn]{aastex631}

\makeatletter
\long\def\frontmatter@title@above{}
\makeatother

\usepackage{graphicx}
\usepackage{amsmath}
\usepackage{url}
\usepackage{xcolor}
\usepackage{tikz}
\usetikzlibrary{calc,fit,positioning,arrows.meta,backgrounds,shapes.geometric}
\definecolor{oiGreen}{HTML}{009E73}
\definecolor{oiOrange}{HTML}{D55E00}
\definecolor{oiBlue}{HTML}{0072B2}
\definecolor{oiGray}{HTML}{4D4D4D}
\definecolor{oiYellow}{HTML}{F0E442}

\usepackage{lettrine}

\newlength{\pixunit}
\newsavebox{\pixelAbox}
\savebox{\pixelAbox}{%
\begin{tikzpicture}[x=\pixunit,y=\pixunit,line width=0.9pt]
 \draw[] (0,-7) -- (2.5,0) -- (5,-7);
 \draw[] (0.95,-2.65) -- (4.05,-2.65);
 \foreach \p in {(0,-7),(2.5,0),(5,-7),(0.95,-2.65),(4.05,-2.65)}
 \fill[] \p circle (0.42);
 \foreach \p in {(0,-7),(2.5,0),(5,-7)}
 \fill[white] \p circle (0.17);
\end{tikzpicture}}

\newcommand{\anchorRangeControls}{1.05--1.2}
\newcommand{\anchorRangeMarkers}{2.6--6.8}

\newcommand{\betaMarkersFull}{47\%}
\newcommand{\bottomTopicName}{instrumentation}
\newcommand{\boundaryShift}{4 percentage points}

\newcommand{\ctrlBetaFull}{51\%}

\newcommand{\declCountsByYear}{26, 38, 153, and 175}
\newcommand{\declRateTwentyFive}{0.81\%}

\newcommand{\deltaEarly}{3.5}
\newcommand{\deltaLate}{1.5}
\newcommand{\disclosureGapFactor}{$\sim$66}

\newcommand{\gammaShiftDown}{4 percentage points}
\newcommand{\gammaShiftUp}{22 percentage points}

\newcommand{\ladderFullRungThree}{56\%}
\newcommand{\ladderFullRungTwo}{61\%}
\newcommand{\ladderFullRungZero}{47\%}

\newcommand{\markerRiseFulltext}{2.6}

\newcommand{\nDeclaredCohort}{392}

\newcommand{\nExpandedHeld}{32}
\newcommand{\nExpandedNew}{30}
\newcommand{\nExpandedWords}{34}
\newcommand{\nFulltext}{207{,}111}

\newcommand{\observedExcess}{2.4}
\newcommand{\piAbstractsCohort}{$75^{+16}_{-16}$\%}
\newcommand{\piAbstractsHeadline}{$61^{+13}_{-12}$\%}
\newcommand{\piAbstractsNuts}{$61^{+13}_{-12}$\%}
\newcommand{\piBottomClass}{44\%}

\newcommand{\piExpandedTracked}{87\%}

\newcommand{\piFulltextHeadline}{$54^{+8}_{-8}$\%}

\newcommand{\piFulltextTwentyFour}{$20^{+3}_{-3}$\%}
\newcommand{\piFulltextTwentySix}{$87^{+8}_{-10}$\%}
\newcommand{\piGridFloor}{36\%}

\newcommand{\piHeadlinePM}{$54^{+8}_{-8}\,(\mathrm{stat},\,95\%)\,^{+26}_{-0}\,(\mathrm{sys,\ background})$}

\newcommand{\piStatOneSigma}{$^{+4}_{-4}$}

\newcommand{\piTopClass}{65\%}
\newcommand{\piUnconstrained}{53\%}
\newcommand{\piWeightShift}{1.7 percentage points}
\newcommand{\piWholeBody}{$52^{+7}_{-7}$\%}
\newcommand{\topTopicName}{cosmology}

\begin{document}

\title{More than half of recent astronomy papers are written with
language-model assistance}

\author[0009-0004-9592-2311]{Serat M. Saad}
\affiliation{Department of Astronomy, The Ohio State University, 140 West 18th
Avenue, Columbus, OH 43210, USA}

\author[0000-0001-5082-9536]{Yuan-Sen Ting}
\affiliation{Department of Astronomy, The Ohio State University, 140 West 18th
Avenue, Columbus, OH 43210, USA}
\affiliation{Center for Cosmology and AstroParticle Physics (CCAPP), The Ohio
State University, Columbus, OH 43210, USA}
\affiliation{Max-Planck-Institut f\"ur Astronomie, K\"onigstuhl 17, D-69117
Heidelberg, Germany}

\begin{abstract}
Language models leave a distinctive vocabulary in the prose they help write, and we measure
how much of the astronomy literature now carries it. From the full text of \nFulltext\
astro-ph papers spanning 2015 to mid-2026, we count those words in each paper and model the
counts, in proportion to paper length, as a mixture of assisted and unassisted writing in a
hierarchical Bayesian model. Papers from before 2020 calibrate the unassisted rate, and the
\nDeclaredCohort\ papers that disclose model use calibrate the assisted one. Our answer depends on
how often these words would appear today if nobody used a model, a rate that must be modeled
rather than observed, so we extend it past 2020 under three assumptions and report all
three. For 2025 that gives \piHeadlinePM\% of papers, the second error being the spread across the three. The estimate stays at or above \piGridFloor\ when we vary that choice, the calibration, and the requirement that adoption only rises. A word list built from the astro-ph
corpus, keeping only words that rose across every subfield, leaves 2025 in the same range. Assisted writing is also getting harder to see, since authors adapt to the
words that reveal it and the marker excess more than halves between 2023 and 2026. Our model
allows for that fading, so it can separate a fainter trace from reduced use. More than half of recent astro-ph papers therefore carry a language-model trace, while
only \declRateTwentyFive\ of 2025 papers disclose it, one declaration for every \disclosureGapFactor\ papers with a trace.
\end{abstract}

\section{Introduction}\label{sec:intro}
After ChatGPT (Chat Generative Pre-trained Transformer) became publicly available in November 2022, a specific set of English words has spread through scientific writing. These words are not technical terms but stylistic choices, like \textit{delve}, \textit{underscore}, \textit{intricate}, \textit{pivotal}, and their relatives, that language models favor and that human writers before November 2022 used rarely. We call them marker words. These models are transformers \citep{Vaswani2017}, scaled to the point where fluent scientific prose is generated at negligible cost \citep{Brown2020}. The pattern was first quantified by \citet{Gray2025}, who estimated from keyword frequencies that at least 1\% of 2023 articles were model-assisted, and by \citet{Liang2024}, who fitted a distributional model to abstracts and found rates of a few percent, rising fastest in computer science.

\citet{Kobak2024} then introduced the approach that most later studies follow. Rather than assume a word list, they measured which words rose anomalously across 14 million PubMed abstracts and concluded that at least 10\% of 2024 biomedical abstracts had been processed with a language model. Extending the method to 1.2 million biomedical full texts, \citet{Holzwarth2026} report that 89\% of papers showed excess model-associated vocabulary by late 2025. Related measurements now exist for peer-review text \citep{Liang2024review}, for condensed-matter physics \citep{Xu2024}, and for preprints across arXiv \citep{Juzek2025, Bao2025}.

The question matters because the documented failure modes of unsupervised model use are concrete, fabricated references above all \citep{Walters2023, Topaz2026}, and because researchers themselves are split on whether such use is acceptable \citep{Kwon2025}. Many journals also ask for disclosure, so it is worth knowing how much goes undisclosed.

Counting excess vocabulary is straightforward. Converting a count into a statement of the form ``$x$\% of papers used a language model'' requires the frequency these words would have had if language models had never been released, and that frequency stopped being observable in November 2022. Statistical precision is not the obstacle, since with hundreds of thousands of papers the observed frequencies are known almost exactly. That frequency can only be supplied by a model, so every estimate in this literature rests on the model chosen for it.

Existing estimators handle that missing frequency by fixing it. The excess-vocabulary bound of \citet{Kobak2024} subtracts a baseline frequency from the observed one and reports the difference as a floor. The extrapolation estimator of \citet{Holzwarth2026} fits the pre-2022 trend of each marker word, extends it linearly, and converts the excess into an estimate that is unbiased if the extension is correct. Neither approach reports uncertainty on the extension itself, and both carry three further restrictions that are worth making explicit.

The first is that each paper is reduced to a binary flag, so a 200-word abstract and a 15{,}000-word monograph count equally. The second is that the background, the rate at which these words appear in writing that no model touched, is one number per year, so no allowance is made for documents growing longer or for topics drifting. The third is that the excess, the factor by which assisted prose lifts that rate, is held fixed, so the method cannot separate more papers using models from models becoming harder to detect.

The literature has already recorded that failure, since \citet{Geng2025} showed that \textit{delve} fell once it became famous in early 2024 while other model-favored words kept rising, so authors adapt and any fixed vocabulary loses power with time. The same group documented the initial shift a year earlier \citep{Geng2024}.

No comparable measurement exists for astronomy, mainly because no clean full-text corpus existed. That changed with the AstroMLab~5 catalog \citep{Ting2025}, which curates 408{,}590 astro-ph papers from 1992 through July 2025 under standardized OCR normalization, together with structured summaries and topic assignments, and makes the underlying full text available on request. We work from an update of that catalog that extends the coverage to August 2026 and holds 433{,}176 papers, of which we analyze the \nFulltext\ first submitted from January 2015 onward that meet our length and topic requirements. We measure both the full text and the abstracts, and we take the full text as our primary evidence. An abstract offers a few hundred tokens, most containing no marker at all, so single abstracts carry little evidence and the abstract fit leans on the whole corpus, while a paper body offers thousands of tokens on its own.

We take up those three restrictions with a hierarchical Bayesian model. We count marker words in each paper against the length of the text we analyze, so that a long paper contributes more evidence than a short one. We calibrate their rate in unassisted writing on the years before 2020, letting it vary with topic and with time, and their rate in assisted writing on the papers whose authors state that they used a model. We also let the difference between those two rates change from year to year, so that a marker word falling out of use as authors adapt to it appears as a fainter trace rather than as less use.

In Section~\ref{sec:methods} we describe the corpus, the vocabulary, the declarations, and the model. In Section~\ref{sec:results} we report the assisted fraction, the fading of the marker excess, a vocabulary drawn from the corpus itself, and the comparison with declared use. In Section~\ref{sec:discussion} we test how far the result depends on our assumptions, place it beside the biomedical measurements, and set out what follows from it.

\section{Methods}\label{sec:methods}

\begin{deluxetable*}{llcl}
\tablecaption{The marker, control, and null vocabularies.\label{tab:baskets}}
\tablehead{\colhead{Basket} & \colhead{Class} & \colhead{$N$} & \colhead{Words}}
\startdata
Basket 1, imported & Strong tells & 18 & \parbox[t]{0.455\textwidth}{\raggedright \textit{boasts, delve, delves, delving, intricacies, intricate, meticulous, meticulously, nuanced, pivotal, realm, realms, showcases, showcasing, tapestry, underscore, underscores, underscoring}} \\[4pt]
Basket 1, imported & Soft tells & 20 & \parbox[t]{0.455\textwidth}{\raggedright \textit{comprehensive, crucial, elucidate, encompass, encompasses, garner, harness, harnessing, holistic, insights, leverage, leveraging, multifaceted, myriad, notably, plethora, robust, seamless, seamlessly, unravel}} \\[4pt]
Basket 2, astronomy-native & Screened & 34 & \parbox[t]{0.455\textwidth}{\raggedright \textit{adhere, align, attribution, categorize, commence, comprehend, comprehension, comprehensive, comprehensively, conversely, disparity, disregard, elucidate, examining, exceeding, excessively, gratitude, grounded, luckily, nonexistent, notably, noteworthy, offering, penultimate, posit, readability, refining, seeming, situated, stood, substantiate, surpass, underscore, utilize}} \\[4pt]
Controls & Neutral & 10 & \parbox[t]{0.455\textwidth}{\raggedright \textit{galaxy, measured, observed, obtained, presented, redshift, sample, spectra, stellar, temperature}} \\[4pt]
Null hedges & Frequency-matched & 15 & \parbox[t]{0.455\textwidth}{\raggedright \textit{adequate, apparent, considerable, consistent, corresponding, derived, distinct, estimated, evident, moderate, notable, reliable, significant, substantial, typical}} \\[4pt]
\enddata
\tablecomments{Basket 1 carries the words imported from the corpus studies of other
fields, separated into strong tells that stay rare in astronomy prose and soft
tells that hold everyday scientific uses. Basket 2 carries the words that the
frozen temporal screen of Section~\ref{sec:discovery} draws from the astronomy
corpus itself, of which \nExpandedNew\ are new relative to Basket 1. The controls anchor the
background rate and the null hedges match the marker frequency distribution
without carrying any tell.}
\end{deluxetable*}

\subsection{Corpus and cohort}\label{sec:m-corpus}

We take all text from the AstroMLab~5 catalog \citep{Ting2025}, in the update described in Section~\ref{sec:intro}. The catalog releases structured summaries and concept annotations, and its OCR full text is available on request. We date every paper by its first submission, decoded from the arXiv identifier, so each paper sits in the era it was first written in. The OCR text corresponds to the latest revision, so a paper posted in 2019 and revised in 2024 can still place post-ChatGPT text in the era we treat as unassisted, and we weigh that leakage in Section~\ref{sec:limits}.

We lowercase the text and tokenize it on contiguous alphabetic strings, so a token is one maximal run of letters, \textit{signal-to-noise} counts three, and mathematics counts none. We remove tables, displayed and inline mathematics, URLs, figure placeholders, and affiliation lines. We then segment the body on section headers into introduction, methods and data, results, discussion, and conclusions, and we analyze only these five section groups. We exclude the abstract so that our full-text analysis does not re-read the same prose, and we exclude
references, acknowledgments, and appendices because their vocabulary is formulaic.

 In many papers we cannot recognize some of the section headers, either because the authors wrote them in an unconventional way or because the OCR drops them. That text holds mostly captions, tables, appendices, and acknowledgements, and it carries the same signal as the named sections, with marker words at a lower rate and the same rise after 2022. We therefore fit the model both ways. When we fit the named sections alone we find \piFulltextHeadline\ for 2025. When we include all the remaining text, which also brings back the 12{,}114 papers that fall below our token threshold on named sections alone, we find \piWholeBody. The two agree, so nothing in our result depends on this choice. We require at least 300 analyzed tokens and a topic assignment, one of the catalog's eight subject classes, which leaves \nFulltext\ papers first submitted between January 2015 and June 2026, with a median of about 2{,}500 analyzed tokens. We run the identical pipeline on the abstracts of these same papers, and we report that measurement in Section~\ref{sec:discussion}. We restrict the abstract fit to the fitted full-text cohort, so the two measurements describe the same papers.

\subsection{Marker, control, and null vocabulary}\label{sec:m-words}

With the text fixed, the measurement needs a vocabulary. We use three word lists, which we call baskets. The model works from the count of each basket in a paper's analyzed text taken as a whole. We also count them inside each of the five sections separately, and those counts enter our checks rather than the model. The marker basket is 38 words that we take from the published biomedical and computer-science lists \citep{Liang2024, Kobak2024, Gray2025}, split between rare, high-specificity words such as \textit{delve} and \textit{underscore} and commoner ones such as \textit{comprehensive} and \textit{notably}.

The control basket is ten neutral astronomy content words, among them \textit{observed}, \textit{galaxy}, and \textit{redshift}, that no language model has any reason to favor. The null basket is fifteen common hedges and intensifiers with pre-2020 frequencies comparable to those of the marker words. Table~\ref{tab:baskets} lists these three baskets in full, along with a second marker basket that we derive from the astro-ph corpus in Section~\ref{sec:discovery}.

We pass the controls and the nulls through the identical pipeline at every stage. These words carry no assisted signal, so any prevalence the pipeline assigns them is a false positive, and we report it as a measure of our own systematic error. We treat this as our main diagnostic.

\subsection{Declarations of model use}\label{sec:m-decl}

The measurement also needs papers that are known to be assisted, and a small number of authors supply them by stating in the paper that they used a language model. We scan every full text for a single sentence that joins a named tool or an explicit language-model phrase, writing or editing vocabulary, and an authorial-use construction such as ``we used'' or ``the authors acknowledge''. We exclude papers about language models themselves by vetoing any paper that mentions models more often than a fixed threshold. We veto a paper when its tool names and model phrases together exceed eight mentions. We also handle two astronomy-specific ambiguities, the Gemini telescopes and colleagues named Claude, by requiring a vendor or version qualifier before we accept an ambiguous name. No such tool existed before 2023, so the roughly 160{,}000 earlier papers test those two ambiguities, and none of them triggers the detector.

We find \nDeclaredCohort\ declarations, none of them before 2023, with counts for 2023 through mid-2026 of \declCountsByYear\ inside the fitted cohort. Declared papers use marker vocabulary at \anchorRangeMarkers\ times the background rate and control vocabulary at \anchorRangeControls\ times it. That contrast is what the fit uses to separate the assisted component from the background. We measure this range against the pre-2020 background, quarter by quarter and smoothed over three, and we use it to anchor the fit of Section~\ref{sec:m-inf}. The corpus-level excess of \observedExcess\ in that section mixes assisted and unassisted papers, and the fitted $e^{\delta_t}$ of Section~\ref{sec:fading} is the excess of assisted prose alone. We re-measure the anchor in every quarter, so the fitted excess can fall below the early declared range as the declared papers themselves grow harder to tell apart.

We audited the detector for missed declarations with a language-model judge. We drew 150 undeclared papers from 2024--2026 at random, screened the closing 3{,}500 characters of each for any loose model-related term, and read the three candidates that surfaced in full.

None is a declaration, since two are funding-program and compute-cluster names that contain the letters AI and one is a paper about language models that the mention-count veto already handles. Zero misses in 150 papers bounds the false-negative rate below 2\% of papers at 95\% confidence. We then read the evidence sentence of every declared paper, all 546 that the scan surfaces corpus-wide. We remove 33 of them, 13 that deny use, which our negation guard now rejects on its own, 17 that discuss language models as a research subject or quote editorial guidance without claiming use, and 3 template statements that say nothing about the writing. We keep the remaining 513, each an authorial statement of use, and almost all of them name grammar, clarity, or language editing as the purpose. We release the removals as a per-paper override table alongside the code. Of the 513, 400 lie in papers with enough analyzed text to enter the feature table, \nDeclaredCohort\ in the fitted cohort, and every count we quote from here on is a cohort count.

\subsection{The mixture model}\label{sec:m-model}

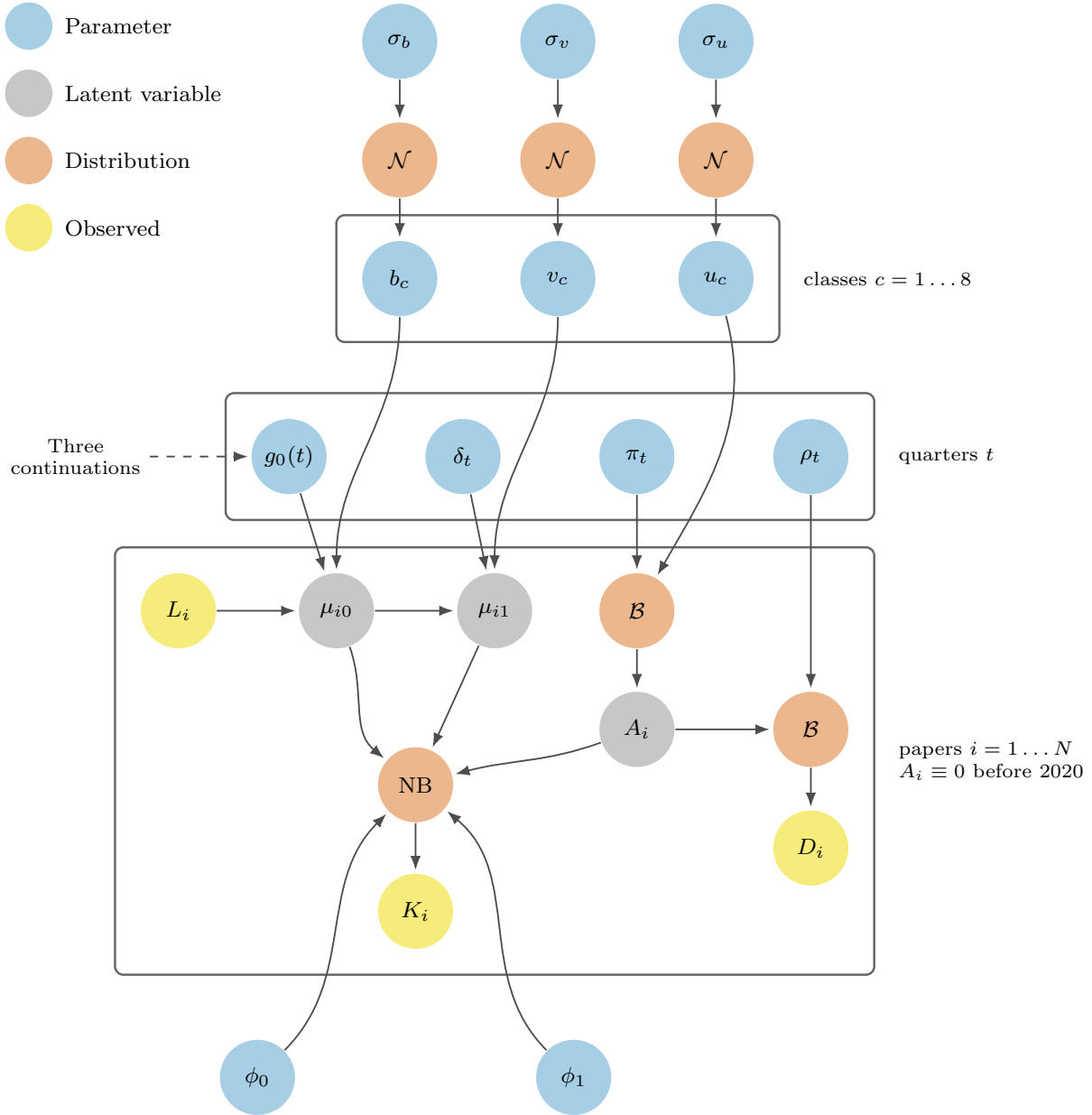
\begin{figure*}[!t]
\centering
\resizebox{0.9\textwidth}{!}{
\begin{tikzpicture}[
    shorten >=1pt, ->, draw=black!70, semithick,
    >={Latex[length=1.8mm]}, font=\footnotesize,
    neuron/.style={circle, minimum size=9.5mm, inner sep=0pt, align=center},
    param/.style={neuron, fill=oiBlue!35},
    inter/.style={neuron, fill=black!22},
    dist/.style={neuron, fill=oiOrange!45},
    obs/.style={neuron, fill=oiYellow!70},
    box/.style={rectangle, draw=black!60, rounded corners=3pt, thick,
        inner sep=9pt},
    lab/.style={font=\scriptsize, text=black},
]

\node[param] (sb) at (2.4, 12.4) {$\sigma_b$};
\node[param] (sv) at (4.4, 12.4) {$\sigma_v$};
\node[param] (su) at (6.4, 12.4) {$\sigma_u$};

\node[dist] (nb_) at (2.4, 10.9) {$\mathcal{N}$};
\node[dist] (nv_) at (4.4, 10.9) {$\mathcal{N}$};
\node[dist] (nu_) at (6.4, 10.9) {$\mathcal{N}$};

\node[param] (bc) at (2.4, 9.4) {$b_c$};
\node[param] (vc) at (4.4, 9.4) {$v_c$};
\node[param] (uc) at (6.4, 9.4) {$u_c$};
\node[box, fit=(bc)(vc)(uc)] (cplate) {};
\node[lab, anchor=west] at ([xshift=5pt]cplate.east) {classes $c=1\ldots8$};

\node[param] (g0)    at (1.0, 7.15)  {$g_0(t)$};
\node[param] (delta) at (3.2, 7.15)  {$\delta_t$};
\node[param] (pi)    at (5.4, 7.15)  {$\pi_t$};
\node[param] (rho)   at (7.6, 7.15)  {$\rho_t$};

\node[box, fit=(g0)(delta)(pi)(rho)] (qplate) {};
\node[lab, anchor=west] at ([xshift=5pt]qplate.east) {quarters $t$};

\node[lab, align=center] (var) at (-1.7, 7.15) {Three\\ continuations};
\draw[dashed] (var) -- (g0);

\node[obs]   (L)   at (-0.4, 5.2) {$L_i$};
\node[inter] (mu0) at (1.6, 5.2)  {$\mu_{i0}$};
\node[inter] (mu1) at (3.6, 5.2)  {$\mu_{i1}$};
\node[dist]  (bern) at (5.4, 5.2) {$\mathcal{B}$};
\node[inter] (A)   at (5.4, 3.7)  {$A_i$};
\node[dist]  (bernD) at (7.6, 3.7) {$\mathcal{B}$};
\node[obs]   (D)   at (7.6, 2.2)  {$D_i$};
\node[dist]  (nbin) at (2.6, 3.0) {NB};
\node[obs]   (K)   at (2.6, 1.4)  {$K_i$};

\node[box, fit=(L)(mu0)(mu1)(bern)(A)(bernD)(D)(nbin)(K)] (pplate) {};
\node[lab, anchor=west, align=left] at ([xshift=5pt]pplate.east)
    {papers $i=1\ldots N$\\ $A_i\equiv0$ before 2020};

\node[param] (ph0) at (0.6, -0.7) {$\phi_0$};
\node[param] (ph1) at (4.6, -0.7) {$\phi_1$};

\draw (sb) -- (nb_);  \draw (sv) -- (nv_);  \draw (su) -- (nu_);
\draw (nb_) -- (bc);  \draw (nv_) -- (vc);  \draw (nu_) -- (uc);

\draw (bc) to[out=-90,in=90] (mu0);
\draw (vc) to[out=-90,in=90] (mu1);
\draw (uc) to[out=-75,in=60] (bern);

\draw (g0) -- (mu0);
\draw (L) -- (mu0);
\draw (mu0) -- (mu1);
\draw (delta) -- (mu1);
\draw (pi) -- (bern);
\draw (bern) -- (A);
\draw (rho) -- (bernD);
\draw (A) -- (bernD);
\draw (bernD) -- (D);

\draw (mu0) to[out=-70,in=140] (nbin);
\draw (mu1) -- (nbin);
\draw (A) to[out=-160,in=15] (nbin);
\draw (nbin) -- (K);
\draw (ph0) to[out=45,in=-135] (nbin);
\draw (ph1) to[out=135,in=-40] (nbin);

\node[anchor=north west, inner sep=0pt] at (-2.6, 12.9) {
    \begin{tikzpicture}[every node/.style={anchor=west}, y=0.85cm]
        \node[param, minimum size=6mm, label={[label distance=4pt]right:Parameter}] at (0,3) {};
        \node[inter, minimum size=6mm, label={[label distance=4pt]right:Latent variable}] at (0,2) {};
        \node[dist, minimum size=6mm, label={[label distance=4pt]right:Distribution}] at (0,1) {};
        \node[obs, minimum size=6mm, label={[label distance=4pt]right:Observed}] at (0,0) {};
    \end{tikzpicture}
};
\end{tikzpicture}}
\caption{The hierarchical Bayesian model, with one plate per replicated index, the eight subject classes, the quarters, and the papers. Each paper contributes its analyzed length $L_i$, its marker count $K_i$, and its declaration flag $D_i$. The background rate $\mu_{i0}$ scales with length and drifts as $g_0(t)$, and assisted papers multiply that rate by the excess $e^{\delta_t}$, which we calibrate on the declared papers. The latent indicator $A_i$ has probability $\pi_t$, the quantity we report, and we fix it to zero before 2020. Distributions appear as their own nodes. The zero-sum class offsets draw from normals with the shared scales $\sigma_b$, $\sigma_v$, and $\sigma_u$, the indicator and the declaration flag draw from Bernoullis, and the count draws from a negative binomial with the dispersions $\phi_0$ and $\phi_1$. The dashed arrow marks where the three background continuations of Section~\ref{sec:m-ident} act, and we do not draw the parameter of Section~\ref{sec:m-robust} that lets declared and silent assisted papers differ in marker rate.\label{fig:model}}
\end{figure*}

\begin{figure*}[!t]
\centering
\includegraphics[width=\textwidth]{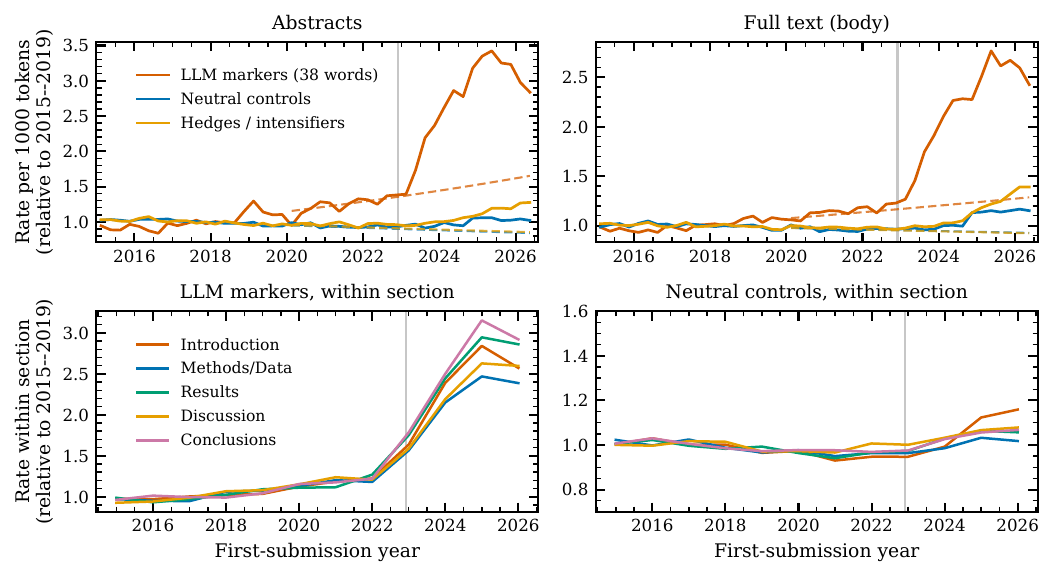}
\caption{We measure the occurrence rate of the three baskets across the decade, in abstracts and in full-text bodies. Top row, occurrence rate per 1000 tokens, normalized to 2015--2019. Dashed segments extend each basket's pre-2020 trend. The markers rise by a factor of \markerRiseFulltext\ in full text by 2025 while the per-token controls stay near flat. Bottom row, the same rates measured inside each section. The markers rise by a similar factor in every section, and the controls drift inside sections by about as much as they drift across the whole document. The grey vertical line in each panel marks the release of ChatGPT.\label{fig:rates}}
\end{figure*}

The corpus gives a length for each paper, the vocabulary gives a marker count, and the declarations give a small set of papers known to be assisted. We combine the three in one model, shown in Figure~\ref{fig:model}. Each paper enters once, with the marker count of its analyzed text taken as a whole, so the five sections are pooled rather than modeled separately. For paper $i$ with analyzed length $L_i$, marker count $K_i$, and latent assistance indicator $A_i$, we write the counts as a two-component mixture of negative binomials,
\begin{align}
K_i \mid A_i{=}0 &\sim \mathrm{NB}(\mu_{i0},\,\phi_0),\nonumber\\
\log\mu_{i0} &= \log L_i + \beta_0 + b_{c_i} + g_0(t_i),\nonumber\\
K_i \mid A_i{=}1 &\sim \mathrm{NB}(\mu_{i1},\,\phi_1),\nonumber\\
\log\mu_{i1} &= \log\mu_{i0} + \delta_{t_i} + v_{c_i},\nonumber\\
A_i &\sim \mathrm{Bern}(\pi_{t_i,c_i}),\nonumber\\
\operatorname{logit}\pi_{t,c} &= m_t + u_c,
\label{eq:model}
\end{align}
and we fix $\pi\equiv0$ for papers submitted before 2020.

Equation~\ref{eq:model} has two levels, and the lower one is easiest to read for a single paper. Paper $i$ has analyzed length $L_i$, subject class $c_i$, and submission quarter $t_i$. If no model touched it, we expect a marker count of $\mu_{i0} = L_i\,e^{\beta_0 + b_{c_i} + g_0(t_i)}$, that is, its length times a rate per token, where $\beta_0$ sets the overall rate, $b_c$ shifts it by subject class, and $g_0$ carries its drift through the decade. If a model did touch it, we expect the larger count $\mu_{i1} = \mu_{i0}\,e^{\delta_{t_i} + v_{c_i}}$, where $e^{\delta_t}$ is the excess that assisted prose carries in quarter $t$ and $v_c$ lets that excess differ by subject class.

Which of the two applies is the latent flag $A_i$, drawn from a Bernoulli distribution with probability $\pi_{t_i,c_i}$. That probability is the quantity we report. We model it on the logit scale so that it stays between zero and one, with $m_t$ giving its rise from quarter to quarter and $u_c$ its offset by subject class. The count we observe, $K_i$, is a negative binomial draw around whichever mean applies, written $\mathrm{NB}(\mu,\phi)$ for mean $\mu$ and dispersion $\phi$, and the two dispersions $\phi_0$ and $\phi_1$ absorb the heavy tail that a Poisson model would misread as structure.

The upper level ties the papers together. Subfields differ in how they write, and they may differ in how quickly they take up models, so each class carries its own background level through $b_c$, its own excess through $v_c$, and its own prevalence through $u_c$. Rather than fit those effects separately for each of the eight subject classes, we draw them from common distributions, so a class with few papers is pulled toward the corpus average instead of swinging the fit on its own. That partial pooling is what makes the model hierarchical.

 We fit all \nFulltext\ papers at once, and every quarter and class informs the others through these shared distributions. The fraction we report is therefore a property of the population rather than a verdict on any one paper, and it is the corpus-wide figure aggregated over the eight classes. We report $\pi_t$ at the class average, since the class offsets sum to zero on the logit scale. When we weight the eight class prevalences by their 2025 paper counts instead, the year figure moves by \piWeightShift. Mixtures of this kind are routine for latent-class problems in astronomy \citep[e.g.,][]{Hogg2010, Foreman-Mackey2014}.

Three features of Equation~\ref{eq:model} answer the three restrictions of Section~\ref{sec:intro} in turn, and together they account for most of the difference from the standard estimator, by which we mean the document-frequency approach of \citet{Kobak2024} and \citet{Holzwarth2026}. The offset $\log L_i$, standard in count regression \citep{Cameron2013}, makes our unit of evidence the count given length rather than the presence of a word anywhere in the document, so a 15{,}000-word paper no longer counts for the same as an abstract. The background terms $b_c$ and $g_0(t)$ let the unassisted rate differ by subject class and drift through the decade, in place of one number per year. Leaving $\delta_t$ free in time lets the excess fade as authors adapt without our model mistaking that fading for a fall in use.

The third of these needs the most care. A free $\delta_t$ can trade against $\pi_t$ quarter by quarter, so it is identified only because the declared papers, whose status we observe, carry information about the excess that the counts alone do not, as Section~\ref{sec:m-ident} describes. That measurement is what lets the excess fade without the estimate of prevalence absorbing the change.

We treat declarations as positive-unlabeled labels \citep{Elkan2008}. A declaration is a positive, but its absence is not a negative, since most assisted authors say nothing. Writing $D_i$ for the declaration flag of paper $i$ and $\rho_t$ for the probability that an assisted author declares, we let a declared paper, with $D_i{=}1$, contribute $\pi\rho\,p_1(K_i)$ and a silent paper, with $D_i{=}0$, contribute $(1-\pi)p_0(K_i)+\pi(1-\rho)p_1(K_i)$, where $p_0$ and $p_1$ are the two negative binomial densities and the quarter and class subscripts are suppressed. Both expressions have the latent $A_i$ already summed out, so the fit involves only continuous parameters.

\subsection{Identification and the missing background}\label{sec:m-ident}

Nothing in Equation~\ref{eq:model} forces its second component to correspond to language-model use. An unconstrained two-component mixture is a flexible curve fit, and given any spread of counts it will find two components, whether or not two processes are present. On the neutral control basket, where the right answer is zero, it does exactly that, as the control fit of Section~\ref{sec:m-robust} shows.

The reason is that prevalence and excess are not separately visible in the counts. The counts show the aggregate lift of marker words above the background, and a given lift can come from many papers each carrying a slight excess or from few papers carrying a strong one. The control basket shows a small lift, which is ordinary vocabulary drift, and a free mixture explains it in the first way.

Our remedy follows the logic of the original spam filter \citep{Sahami1998}. Rather than let the counts alone define the assisted component, we anchor it to the papers we know are assisted. Their declaration flag is observed, so they enter the same likelihood as known positives, and the prior on $\delta_t$ is centered on the excess their counts display. This is not a calibration carried out before the fit, since the excess and the prevalence are estimated together.

Holding $\delta_t$ near the declared value rules out the first reading, and the model can no longer turn ordinary drift in control words into prevalence. We list every prior in Table~\ref{tab:priors}. We let the background drift $g_0$ run quadratic in time, with the linear term free at $\mathcal{N}(0,0.5)$ per quarter-scale unit and the curvature confined to $\mathcal{N}(0,0.05)$, and the pre-2020 papers pin it. We give the log excess $\eta_t$ a random walk pulled to the declared anchor with width 0.2 in each quarter, so we smooth it rather than leave it free, and the anchor floor at 1.05 keeps the two components from merging, though we do not constrain $\delta_t$ positive. We impose rising adoption in the primary specification as a hard constraint, softplus-positive steps on the prevalence walk, and we remove it in the unconstrained variant.

\begin{deluxetable*}{lll}
\tablecaption{Priors of the hierarchical model. All class effects sum to
zero across the eight classes.\label{tab:priors}}
\tablehead{\colhead{Parameter} & \colhead{Role} & \colhead{Prior}}
\startdata
$\beta_0$ & log background rate per token & $\mathcal{N}(-7,3)$ \\
$b_c$ & class background offsets & $\mathcal{N}(0,\sigma_b)$, $\sigma_b\sim\mathcal{N}^+(1)$ \\
$g_0$ slope, curve & background drift & $\mathcal{N}(0,0.5)$, $\mathcal{N}(0,0.05)$ \\
$\phi_0,\ \phi_1$ & overdispersions & $\mathcal{N}^+(5)$ \\
$\eta_t=\log\delta_t$ & assisted log excess & walk, steps $\mathcal{N}(0,\sigma_\eta)$, $\sigma_\eta\sim\mathcal{N}^+(0.4)$; pulled to anchor, width 0.2 \\
$v_c$ & class excess offsets & $\mathcal{N}(0,\sigma_v)$, $\sigma_v\sim\mathcal{N}^+(0.3)$ \\
$m_0$ & initial logit prevalence & $\mathcal{N}(-4,2)$ \\
$\Delta m_t$ & prevalence steps & softplus-positive, $\mathcal{N}^+(1)\times\sigma_m$, $\sigma_m\sim\mathcal{N}^+(0.6)$ \\
$u_c$ & class prevalence offsets & $\mathcal{N}(0,\sigma_u)$, $\sigma_u\sim\mathcal{N}^+(0.5)$ \\
$\rho_a,\ \rho_b$ & declaration propensity & $\mathcal{N}(-4,2)$, $\mathcal{N}(0,1)$ \\
\enddata
\end{deluxetable*}

Three other quantities come from the data rather than from assumption. We set $\pi\equiv0$ before 2020, so the 106{,}000 papers from those years act as known negatives, and the background level, its scaling with length, its variation by topic, and its scatter are estimated from them inside the same fit rather than set beforehand. The neutral controls bound how fast ordinary astronomy vocabulary drifts. The declaration counts by year give $\rho_t$.

A fourth quantity is assumed rather than measured. We impose a monotonicity constraint, requiring the assisted fraction not to fall from one quarter to the next, which is a statement about how adoption behaves rather than anything the counts show, and we report every number without it as well.

One quantity none of this fixes is the background itself after 2020. The background is the rate at which human authors would have used these words anyway, and it does not stand still. Scientific vocabulary drifts, the mix of subfields changes, and the share of authors writing in a second language grows. Before 2020 we can watch that drift directly. After 2020 we cannot separate it from model use, so it has to be assumed rather than measured, and rather than adopt one continuation silently, we fit the model in three variants.

The three variants share Equation~\ref{eq:model} and differ only in the form taken by $g_0$ after 2020. In the \textit{linear} variant we continue the slope measured before then, so that $g_0(t) = g_0(2020) + \kappa\,(t-2020)$ with $\kappa$ the pre-2020 trend, which for these words was already upward and which is what the standard estimator assumes. In the \textit{frozen} variant we set $g_0(t) = g_0(2020)$, so we attribute every later change to assistance, and this gives the highest answer. In the \textit{tracked} variant we set $g_0(t) - g_0(2020)$ to the measured drift of the neutral controls, and this falls between the other two. The tracked variant assumes that the control drift carries no assistance of its own, which is why we report it as one of three rather than adopt it.

We take the linear variant as our primary specification, since it is the most conservative of the three. We keep the spread between the three separate from the statistical uncertainty of any single fit, and report the other assumptions we vary alongside it rather than folding them in.

\subsection{Inference and calibration}\label{sec:m-inf}

With the model specified and anchored, we fit it in two stages. The model carries one parameter per quarter for each of the background drift, the excess, and the prevalence, along with the class effects and the two dispersions, and the variants of Section~\ref{sec:m-robust} require the whole fit to be repeated dozens of times. For that grid we maximize the log posterior with every bounded parameter transformed onto an unbounded scale and take uncertainties from the curvature at the mode, the Laplace approximation \citep[cf.][]{Gelman2013}, which converges in about a minute per variant.

We then sample the primary specification in full with Hamiltonian Monte Carlo \citep{Neal2011, Hoffman2014, Betancourt2017}, running four chains of 1{,}000 draws after 1{,}000 warmup steps and starting every chain at the mode the first stage found. We judge convergence with the two standard diagnostics, the rank-normalized $\widehat{R}$, which compares the spread within each chain against the spread across chains, and the effective sample size, which counts how many independent draws the correlated ones are worth \citep{Vehtari2021}. Every parameter returns $\widehat{R}$ of 1.00 with effective sample sizes between 1{,}300 and 6{,}100, and we quote every primary number in this paper from the sampled posterior. The mode-based approximation reads high in this model, by 13 percentage points on the 2025 fraction, because the prevalence walk is skewed and a Gaussian at the mode overweights its upper tail. The abstract fit repeats the pattern, with \piAbstractsHeadline\ at the mode and \piAbstractsNuts\ from the sampler. We sample the three background continuations and the unconstrained walk the same way. The remaining variants stay mode-based, and we read their shifts about the sampled primary rather than as levels.

We also test the estimator by injection and recovery. We build synthetic corpora from the real pre-2020 papers, so that lengths, topics, and dispersion are realistic by construction. In each corpus we mark a known fraction of the papers as assisted, raise the marker rate of those papers by a known factor, redraw their counts, and then run the estimator to see what fraction it returns. We run the test twice. In the first pass we put no declared papers in the synthetic corpora and run the estimator unanchored, which exposes the mixture arithmetic on its own. In the second pass we mark declared papers inside the simulation at the observed declaration rate and fit with the anchor in force, exactly as in the real analysis. In both passes we redraw counts from the fitted negative binomial as a gamma--Poisson mixture, hold the background at its true value, apply the injected excess as one factor across classes and quarters, and inject prevalences from 5\% to 75\%.

At an injected excess of 2.5 or more, both passes recover the injected fraction within two or three percentage points across the whole grid, and the corpus-level excess of \observedExcess\ that we measure places most of our fit in that regime. At an excess of 1.5, the value the fitted excess reaches by 2026, the two passes separate. The unanchored recovery scatters, which is the prevalence and excess trade of Section~\ref{sec:m-ident} reappearing once the components sit close together. The anchored recovery holds, with the recovered mean within six percentage points of the injected truth and the scatter within eight for every prevalence of 15\% and above, at a tenth of the corpus size. The one cell that stays open, a prevalence near 5\% with an excess of 1.5, leaves too few declared papers in the simulation to form an anchor, and no quarter the data prefer sits there. The late quarters therefore rest on the declared calibration, and the calibration passes the test with the anchor it actually uses.


\section{Results}\label{sec:results}

\begin{figure*}[!t]
\centering
\includegraphics[width=\textwidth]{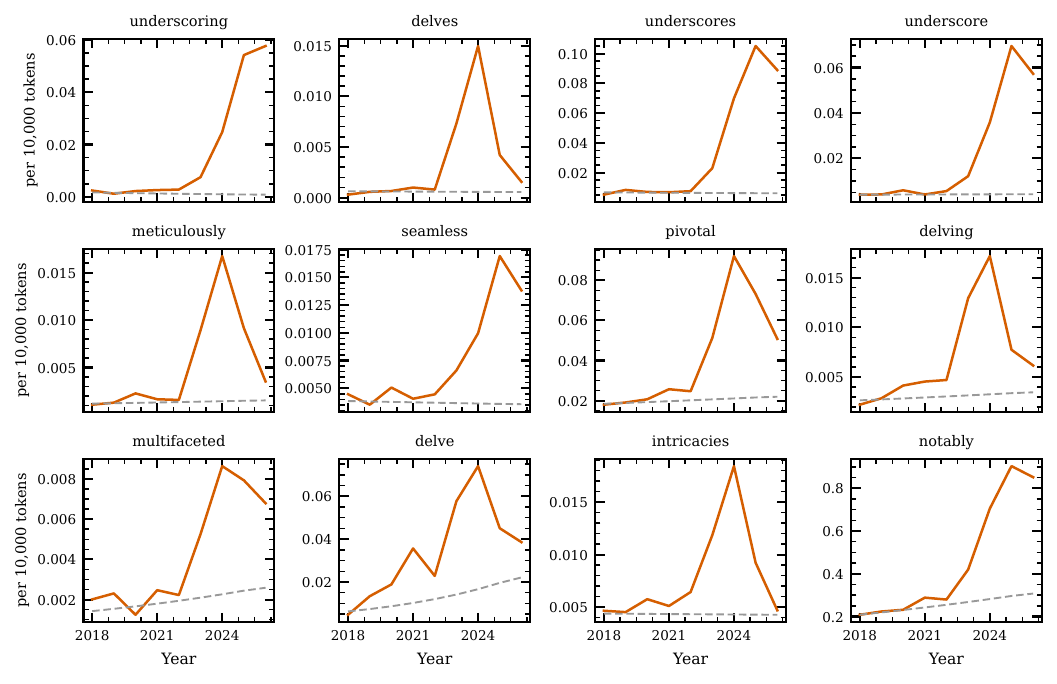}
\caption{We follow each marker word separately to show that the aggregate signal is steadier than any of its parts. The orange curve is the observed occurrence rate of each of the twelve markers with the largest recent excess, and the grey dashed curve is the background that we extrapolate from before 2020. Each curve is the word's rate in assisted prose, recovered by undoing the mixture with the whole-basket prevalence, so no word chooses the papers by which it is measured. The words do not move together, since \textit{delve} peaks in 2024 and returns toward its baseline while \textit{underscore} and \textit{notably} are still rising in 2026. An estimator tied to a fixed word list loses power as the words themselves change. In the model fitted here the excess is free to change from year to year and is held in each period to the papers whose authors declare model use.\label{fig:words}}
\end{figure*}

\begin{figure}[!t]
\centering
\includegraphics[width=\columnwidth]{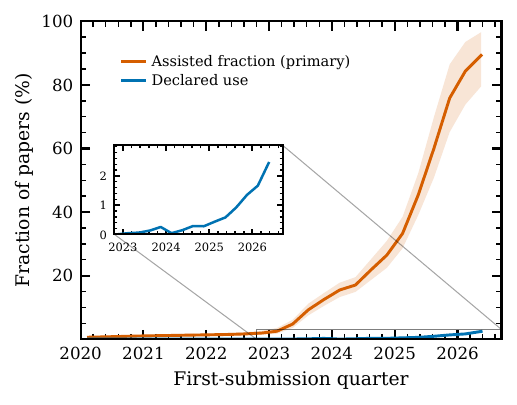}
\caption{The assisted fraction of astro-ph papers by first-submission quarter, under the primary specification, linear background drift, with its 95\% credible interval. The fraction passes half during 2025. The background variants of Section~\ref{sec:m-ident} shift this curve, and Section~\ref{sec:m-robust} reports the shifts. The lower curve is the fraction of papers whose authors declare model use, on the same scale, and the inset enlarges it. \label{fig:results}}
\end{figure}

\subsection{The assisted fraction}\label{sec:prevalence}

Our main result appears in Figure~\ref{fig:results}. Under the primary specification, the fraction of astro-ph papers whose analyzed prose shows a language-model trace reaches \piHeadlinePM\% in 2025, up from \piFulltextTwentyFour\ in 2024, and it reaches \piFulltextTwentySix\ in the first half of 2026. Errors written as plus and minus are 95\% credible intervals of a single fit, here and throughout, and they carry statistical uncertainty alone. The second term is the spread across the three background continuations, one-sided because the linear one is the most conservative. It is not a full systematic budget, since the other assumptions we vary in Section~\ref{sec:m-robust} move the estimate separately and we report them there. At one sigma the 2025 interval is \piStatOneSigma\%.

The linear continuation sets the floor and the frozen background the ceiling, and the control-tracked variant falls between them. The ordering is itself informative. The marker words were already drifting upward before 2020, so extrapolating that drift absorbs part of the post-2022 excess and gives the most conservative answer. The frozen and tracked variants instead require ordinary vocabulary drift to have stopped in 2020, which the pre-2020 record gives no reason to expect, and this is why we take the linear continuation as the primary. When we remove the monotonicity constraint, which forbids the assisted fraction from falling between one quarter and the next, the 2025 figure moves by less than a percentage point, to \piUnconstrained, and the two trajectories track each other through the whole ramp. The largest single sensitivity sits instead in the calibration, the declared-to-silent ratio of Section~\ref{sec:m-robust}.

The estimand we adopt is narrow, the fraction of papers whose analyzed prose is more consistent with our calibrated assisted component than with the background. This is weaker than the claim that a machine wrote the paper and stronger than the observation that a paper contains one suspicious word. We count an author who ran a single section through a language model for grammar. We do not count model use that leaves no lexical trace, whether because the author edited it away or because the help was with code or ideas rather than prose. Every number we report is a floor in that sense. We use the per-paper probabilities only as weights inside the aggregate, and we do not publish them.

\begin{figure}[!t]
\centering
\includegraphics[width=\columnwidth]{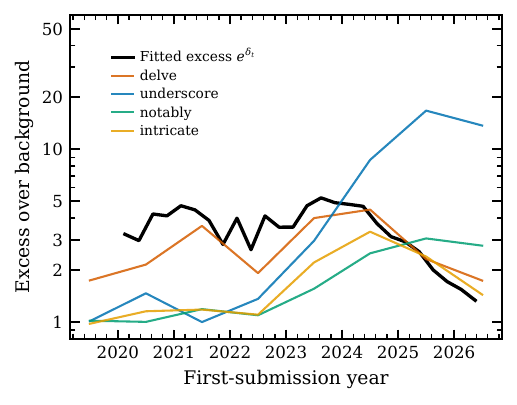}
\caption{The excess of assisted prose fades as adoption rises. The black curve is the fitted excess $e^{\delta_t}$, the factor by which assisted papers lift their marker rate above background, and it falls from \deltaEarly\ in early 2023 to \deltaLate\ by 2026. The colored curves follow four marker words, each deconvolved with the whole-basket prevalence so that no word selects the papers by which we measure it. \textit{Delve} falls after mid-2024 while \textit{underscore} continues to rise, and the words do not move together, so a fixed word list loses power with time while the fitted excess absorbs the change.\label{fig:fading}}
\end{figure}

\subsection{Time evolution of the marker excess}\label{sec:fading}

Figure~\ref{fig:fading} shows our second result. The fitted excess of assisted prose falls from \deltaEarly\ times the background in 2023 to \deltaLate\ by 2026 while the prevalence rises. Taken together, these two trends are the population-level form of the adjustment that \citet{Geng2025} measured word by word. More papers pass through a language model while each leaves a fainter lexical trace.

In Figure~\ref{fig:words} we show the same result one word at a time. The rate we observe for a word blends assisted and unassisted papers in the proportion the model estimates, so we undo that blend, removing the unassisted contribution and dividing by the fitted prevalence, and each curve then shows the word's rate in assisted prose. The prevalence we use comes from the whole basket rather than from the word in question, so no word chooses the papers by which it is measured.

\textit{Delve} falls after mid-2024, once it became a recognized marker of model use, while \textit{underscore} and \textit{notably} continue to rise. The implication is that fixed word lists lose their power within a few years, and any estimator that holds the excess constant will mistake the fading of markers for a fall in use. In our model the excess is free to change and is held in each period to the declared papers, so the same motion registers as fading.

\begin{figure*}[!t]
\centering
\includegraphics[width=\textwidth]{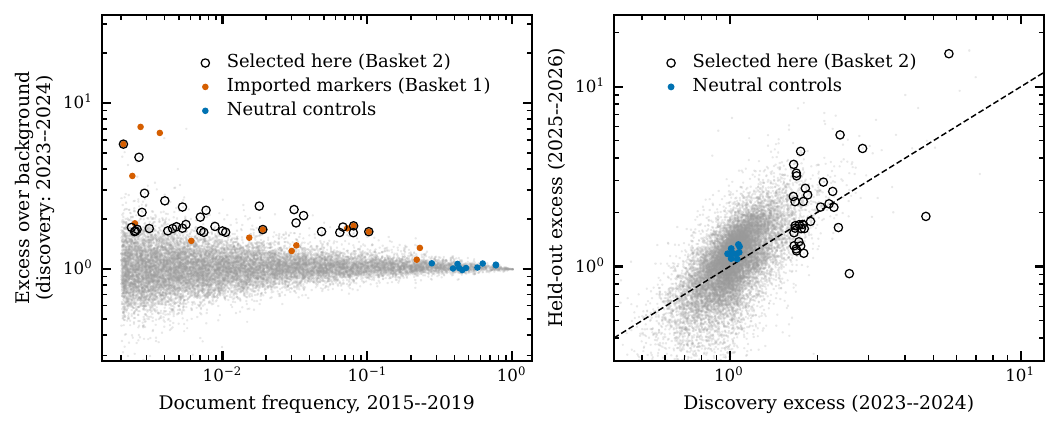}
\caption{We derive a marker vocabulary from the astro-ph corpus and test it on years the selection never saw. Left, every word with a stable pre-2020 background, placed by its 2015--2019 document frequency and its 2023--2024 excess over the extrapolated background. Circled words pass our selection screen, which requires a stable background, a rise in the discovery years, and a spread across all eight subject classes. The last criterion separates writing style from subfield vocabulary. Right, the same selection, fixed before these years were examined, evaluated on 2025--2026. The selected words scatter about the one-to-one line while the neutral controls sit at unit excess on both axes. The filled points in the left panel are the 38 imported words of basket 1. The screen selects four of them as well, and a word in both baskets appears as a ringed point.\label{fig:discovery}}
\end{figure*}

\subsection{Marker vocabulary derived from the corpus}\label{sec:discovery}

The 38-word basket of Section~\ref{sec:m-words} comes from studies of biomedical and computer-science writing \citep{Liang2024, Kobak2024, Gray2025}, and language models may mark astronomy prose differently. We therefore derive a second basket from the astro-ph corpus, using a temporal split that we fix in advance. We draw the candidates from the catalog's companion $n$-gram index, which supplies word frequencies by year across the corpus and is browsable online.\footnote{\url{https://tingyuansen.github.io/astro-ph_ngram_viewer/}} We require a candidate to occur in at least 0.2\% of pre-2020 papers with a stable background, to rise by at least 60\% over the extrapolated background in the discovery years 2023 and 2024, and to spread across all eight subject classes. The last requirement separates writing style, which appears in every subfield at once, from science, which does not. We use the catalog's own eight subject classes, which sort every paper into cosmology and nongalactic physics, galaxy physics, high-energy astrophysics, solar and stellar physics, earth and planetary science, numerical simulation, instrumental design, or statistics and machine learning. We find the 2025 prevalence runs from \piBottomClass\ in \bottomTopicName\ to \piTopClass\ in \topTopicName.

The screen promotes \textit{noteworthy} and \textit{comprehensively}, and it rejects the pulsar-timing vocabulary that rose after the 2023 gravitational-wave background detections. We remove author-name fragments and facility names by requiring each candidate to appear in an English dictionary and by excluding a short list of names by hand. The selected words appear in Table~\ref{tab:baskets}.

Figure~\ref{fig:discovery} shows the selection and its test on later years. The basket we fix in advance has \nExpandedWords\ words, of which \nExpandedNew\ are new relative to the imported basket. Of the \nExpandedWords, \nExpandedHeld\ independently show more than 20\% excess in 2025 and 2026, years that our selection never used, and the neutral controls, whose excess we measure the same way over those two years, sit between 0.98 and 1.08.

When we fit the model with this basket we obtain $80^{+10}_{-12}$\% for 2025 under linear drift and \piExpandedTracked\ under tracked drift, both inside the range the imported words give. The linear figure exceeds the $67^{+12}_{-11}$\% that the imported basket gives under the same background, which is what we would expect if a vocabulary drawn from astronomy prose catches uses that a borrowed list misses. On that reading the imported basket understates the fraction.

\subsection{Comparison with declared use}\label{sec:gap}

Figure~\ref{fig:results} sets our estimate beside declared use on the same scale. In 2025, \declRateTwentyFive\ of full texts contain a declaration against an estimated prevalence of \piFulltextHeadline, so roughly \disclosureGapFactor\ papers carry a language-model trace for every one that says so.

The gap is not an artifact of the detector. The audit of Section~\ref{sec:m-decl} found no missed declaration in 150 undeclared papers, which bounds the false-negative rate below 2\%, so nearly all of the silence is real rather than overlooked. Nor is the gap closing. Declarations grew by a factor of four between 2023 and 2025 while our estimated prevalence grew faster, and the two curves separate across the whole period.

The declarations themselves are narrow in scope. Almost every declaration we read names grammar, clarity, or language editing as its purpose, which is also the use our estimand is built to catch.

\section{Discussion and conclusions}\label{sec:discussion}

We find that more than half of recent astro-ph papers carry a language-model trace in their prose. Our 2025 fraction is \piFulltextHeadline\ under the primary specification, and it never falls below \piGridFloor\ across the variants we ran. By the first half of 2026 the fraction reaches \piFulltextTwentySix, at the 89\% that \citet{Holzwarth2026} report for biomedical full texts in late 2025. No astronomy-specific estimate existed before this work.

\subsection{Robustness of the estimate}\label{sec:m-robust}

We read these fractions with their definition in mind. We count a paper as assisted when its analyzed prose shows the marker vocabulary at the strength we see in the declared papers, and almost all declarations name grammar, clarity, or language editing as the purpose, so we estimate the fraction of papers in which a model played some part in the writing, not the fraction written by a model. That floor does not rest on the hierarchy. On the per-token scale the markers rise by a factor of \markerRiseFulltext\ by 2025, and dividing that excess by the excess we measure in declared papers already puts the assisted fraction at \ladderFullRungTwo, above half, in a single line of arithmetic. The same ratio in 2026, where we measure the corpus near 2.5 times its background and the declared papers near 3 times theirs, lands within a few points of the sampled fraction. Nor does it rest on how the OCR divides a paper, since the markers rise by a similar factor inside each of the five named sections, as the lower panels of Figure~\ref{fig:rates} show, while the controls drift inside sections by about as much as they drift across the whole document.

The model exists to control the assumptions that arithmetic leaves implicit, and each of them can be tested in turn. Our model gives \piFulltextHeadline\ for 2025. Holding the excess fixed in time, as the standard estimator does, lowers that to \ladderFullRungThree. Flattening the background as well raises it to \ladderFullRungTwo, since ordinary vocabulary drift is then read as assistance instead of being absorbed. Asking only whether a marker word appears in a paper, rather than how often it appears per token, lowers it to \ladderFullRungZero. That last assumption costs the most. Requiring adoption not to fall from quarter to quarter costs nothing in the sampled fit, since Section~\ref{sec:prevalence} gives \piUnconstrained\ without it. Moving the known-negative boundary, the date before which every paper is treated as unassisted, from 2020 to the quarter before ChatGPT was released changes the estimate by \boundaryShift.

One assumption matters more than the rest, because it concerns the papers the whole calibration rests on. The declaration counts give $\rho_t$, the probability that an assisted author declares, but they say nothing about whether those authors use marker words at the same rate as assisted authors who stay silent. We assume that they do, since the silent papers are never identified, and we fix the ratio at one in the primary specification. Setting it to two, declared authors twice as marker-heavy as their silent peers, raises the 2025 estimate by \gammaShiftUp, and setting it to one half lowers it by \gammaShiftDown.

When we run the same model on the abstracts of these same papers, restricting the fit to the full-text cohort, we find \piAbstractsCohort\ for 2025, consistent with the full-text range. Abstracts remain the weaker instrument, and our intervals from them are roughly twice as wide because each paper offers thirty times fewer tokens.

\subsection{Why document-frequency estimates come out lower}\label{sec:m-compare}

Our result stands apart from what the standard estimator returns on the same corpus. The extrapolation estimator of \citet{Holzwarth2026}, run on the 38 markers on full text, gives a 2025 assisted fraction of \betaMarkersFull. The estimator was built for abstracts, where a document is a few hundred tokens, and what fails here is the transfer to documents thirty times longer, not the arithmetic itself. On full text it does not pass our null test. When we apply the same arithmetic to the neutral control basket we obtain \ctrlBetaFull, and when we apply it to the null hedges we obtain a similar number. Words with no possible connection to language models therefore yield an apparent prevalence as large as the marker signal.

The mechanism is document length, since the median body grew from about 4{,}500 tokens in 2015 to 6{,}400 in 2025 and the probability that a fixed-rate word appears at least once rises with length. A statistic that asks only whether a word appears therefore treats longer papers as more contaminated. Measuring occurrences per token instead removes most of the control excess, as the top row of Figure~\ref{fig:rates} shows, and it is why our own unit of evidence is the count given length. The same effect explains the apparent order-of-magnitude disagreement between abstract-based and full-text estimates in the earlier literature. Once we measure counts per token and calibrate the assisted component separately in each corpus, the two agree, as Section~\ref{sec:m-robust} shows.

A different failure appears in our own model when its anchor is removed. Without the anchor the mixture assigns $57^{+5}_{-4}$\% of papers to an assisted component built from \textit{galaxy}, \textit{redshift} and their like, a fraction invented for words with no link to model use. With the anchor, declared papers use control words at only \anchorRangeControls\ times the background, so no control prevalence is identified at all.

The control counts retain only ordinary vocabulary drift, an order of magnitude below the marker signal, and the tracked-drift variant folds that drift into the background, so it never enters the headline.

\subsection{Limitations}\label{sec:limits}

Several limits are specific to this work. Our estimand is a trace in analyzed prose, so we cannot see use that leaves no lexical trace, and every number we report is a floor. We work from OCR text of the latest revision of each paper, so the era we treat as unassisted can hold a small admixture of post-2022 revisions. We also calibrate the assisted component on the declared papers, who are a selected minority, so the sweep of Section~\ref{sec:m-robust} stands in for a comparison we cannot make, the marker rate of authors who used a model and said nothing.

A further limit applies to every method in this literature, ours included. When the assisted and background components sit close together, a small error in the assumed background becomes a large error in the fraction. For vocabulary with a strong excess, such as our markers, this matters little, but for vocabulary with a weak excess it removes the answer and more data does not help. This is why we report the null test in units of excess rate rather than as a prevalence, and why we would not trust a prevalence built on low-excess words.

A final limit is that the unassisted rate we call the background may itself carry language-model influence. Some of the drift the controls retain may be assistance our detector misses, since authors who use a language model absorb its habits and leave the same faint trace in ordinary words. Some may be diffusion, since authors read a great deal of model-written prose and may take up its vocabulary without ever running their own manuscript through anything. Part of the rise in the controls, and more of the larger rise in the hedges, may therefore be model-driven. Both rises stay small beside the marker rise, and the drift variants fold them into the background, which is why the held-out screen of Section~\ref{sec:discovery} finds the controls at unit excess against their trend.

That possibility is why we bracket the background rather than tie it to the controls, and it works in our favour. Words that carry a faint trace but sit in the control or null baskets, among them the discourse connectors that language models also favor, are absorbed into the background instead of being counted as signal, so each one lowers the fraction we report.

\subsection{Concluding remarks}\label{sec:follows}

We do not interpret our result as misconduct at scale. The lexical shift is strongest in the writing of non-native English speakers, where others have interpreted it as a leveling of linguistic barriers rather than as misconduct \citep{Lin2025}, and much of what we detect is likely the grammar and style assistance that journal policies already permit. At the prevalence we measure, that use is no longer unusual, and a stigma attached to it is harder to justify than the practice itself.

Enforcement is the half of the problem that cannot be won. The excess falls from \deltaEarly\ to \deltaLate\ within three years while use rises, so a policy that depends on catching offenders is chasing a signal its targets are already erasing. Stronger instruments have been proposed, watermarking among them, which is a deliberate version of the statistical trace we exploit here. Whether such measures are desirable is a live debate, but any of them strong enough to work would have to be imposed on the whole literature in order to police a practice that most of the literature already follows, and would be invasive in proportion.

Disclosure is the half that can be won. We find roughly \disclosureGapFactor\ papers carrying a trace for every one that declares, and the declarations we do find name grammar, clarity and language editing, which is what journal policies already allow. Authors are therefore not concealing misconduct so much as avoiding a stigma, and researchers remain divided on whether the use is acceptable at all \citep{Kwon2025}. Removing that stigma costs nothing, and it would turn a norm most of the field ignores into one most of the field has no reason to break. It is also the more urgent of the two, since the trace we rely on here is fading and the measurement will not stay available indefinitely.

\section*{Acknowledgments}

SMS is supported by the Distinguished University Fellowship awarded by The Ohio State University. YST is supported by NSF under Grant AST-2406729 and by a Humboldt Research Award from the Alexander von Humboldt Foundation.

We built this work on the AstroMLab~5 astro-ph catalog and on the companion $n$-gram index that supplies the word frequencies of Section~\ref{sec:discovery}.

Given the subject of this paper, we state our own practice plainly. We used Claude Code for code development. We drafted and revised the text ourselves, then used Claude Opus 5 to edit the prose and grammar throughout.

\section*{Data and code availability}
The corpus is the AstroMLab~5 catalog \citep{Ting2025}, in an update that extends its coverage from July 2025 to August 2026, and its OCR full text is available on request. We do not redistribute it. We release the analysis code and the derived data at \url{https://github.com/seratsaad/llm-in-astro-ph}. The release holds the per-paper feature tables with word counts, section lengths, and declaration flags, the frozen marker, control, and null baskets, the declaration table with every evidence sentence and the audited override list, the injection-recovery grids, and the posterior of every specification in the variant grid alongside the sampled primary. We do not release a per-paper assisted probability, since we calibrate the model for population inference rather than for judging individual papers.

\software{NumPy \citep{2020Natur.585..357H}, SciPy
\citep{2020NaMet..17..261V}, Matplotlib \citep{2007CSE.....9...90H}, pandas
\citep{2020zndo...3509134R}, JAX \citep{jax2018github}, PyMC
\citep{2023PeerJCS...9.1516A}, ArviZ \citep{2019JOSS....4.1143K}}

\bibliographystyle{mnras}
\bibliography{refs}

\end{document}